\documentclass[10pt,conf]{IEEEtran}
\ifCLASSINFOpdf \else \fi
\usepackage{geometry}
\usepackage[normalem]{ulem}
\usepackage{multirow}
\usepackage{bbold}
\usepackage{ifpdf}
\usepackage{ulem} 
\usepackage{subcaption}%
\usepackage[cmex10]{amsmath}
\usepackage{amssymb}
\usepackage{verbatim}
\usepackage{algorithm}
\usepackage{algorithmic}
\usepackage{array}
\usepackage{latexsym}
\usepackage{color}
\usepackage{textgreek}
\usepackage{subscript}
\usepackage{rotating}
\usepackage{booktabs,multirow}
\usepackage{mdframed}	
\usepackage{tabularx}
\usepackage{amsmath}
\usepackage{makecell}
\usepackage{tabto}
\usepackage{mathrsfs}
\usepackage{url}
\usepackage{cite}
\usepackage{listings}
\usepackage{array}
\usepackage[table]{xcolor}
\usepackage{ragged2e}
\usepackage{booktabs}

\usepackage{graphicx}
\usepackage{comment}
\usepackage[cmex10]{amsmath}
\usepackage{amssymb}
\usepackage{bbold}
\usepackage{float}
\usepackage[cmex10]{amsmath}
\usepackage{amssymb}
\usepackage{verbatim}
\usepackage{array}
\usepackage{latexsym}
\usepackage{color}
\usepackage{tabularx} 
\usepackage{textgreek}
\usepackage{subscript}
\usepackage{rotating}
\usepackage{booktabs,multirow}
\usepackage{mdframed}	
\usepackage{tabularx}
\usepackage{amsmath}
\usepackage{makecell}
\usepackage{tabto}
\usepackage{mathrsfs}
\usepackage{url}
\usepackage{cite}
\usepackage{listings}
\usepackage{array}
\usepackage[table]{xcolor}

\definecolor{pinkpurple}{rgb}{0.6, 0.1, 0.9} 
\usepackage{algorithmic}
\usepackage{array}
\usepackage{soul}
\sethlcolor{green}
\usepackage{url}
\usepackage{pifont}
\usepackage{xcolor}
\usepackage{multirow}
\usepackage{graphicx} 

\begin{document}


\title{Integrating Language Models for Enhanced Network State Monitoring in DRL-Based SFC Provisioning}

\author{
\IEEEauthorblockN{Parisa~Fard~Moshiri$^1$, Murat Arda Onsu$^1$, Poonam Lohan$^1$, Burak Kantarci$^1$, Emil Janulewicz$^2$,}\\
\IEEEauthorblockA{\textit{$^1$University of Ottawa, Ottawa, ON, Canada}\\
\textit{$^2$Ciena, 383 Terry Fox Dr,
Kanata, ON K2K 2P5, Canada}\\
$^1$\{parisa.fard.moshiri, monsu022, ppoonam, burak.kantarci\}@uottawa.ca,~$^2$ejanulew@ciena.com}
\vspace{-0.2in}}

\maketitle
\thispagestyle{empty}
\pagestyle{empty}
\begin{abstract}

Efficient Service Function Chain (SFC) provisioning and Virtual Network Function (VNF) placement are critical for enhancing network performance in modern architectures such as Software-Defined Networking (SDN) and Network Function Virtualization (NFV). While Deep Reinforcement Learning (DRL) aids decision-making in dynamic network environments, its reliance on structured inputs and predefined rules limits adaptability in unforeseen scenarios. Additionally, incorrect actions by a DRL agent may require numerous training iterations to correct, potentially reinforcing suboptimal policies and degrading performance. This paper integrates DRL with Language Models (LMs), specifically Bidirectional Encoder Representations from Transformers (BERT) and DistilBERT, to enhance network management. By feeding final VNF allocations from DRL into the LM, the system can process and respond to queries related to SFCs, DCs, and VNFs, enabling real-time insights into resource utilization, bottleneck detection, and future demand planning. The LMs are fine-tuned to our domain-specific dataset using Low-Rank Adaptation (LoRA). Results show that BERT outperforms DistilBERT with a lower test loss (0.28 compared to 0.36) and higher confidence (0.83 compared to 0.74), though BERT requires approximately 46\% more processing time.
\end{abstract}


%
\IEEEpeerreviewmaketitle
\vspace{-2mm}
\section{Introduction}
Service Function Chain (SFC) provisioning and Virtual Network Function (VNF) placement are critical for optimizing network performance in modern architectures like Software-Defined Networking (SDN) and Network Function Virtualization (NFV). NFV improves agility while decreasing operational and capital expenditures by using virtualization to separate software from physical infrastructure and place network services on general purpose hardware, such as data centers (DCs) \cite{KAUR2020}. SFC enhances NFV's advantages by sequencing VNFs to enable services such as Cloud Gaming (CG), Augmented Reality (AR), VoIP (Voice over IP), Video Streaming (VS), Massive IoT (MIoT), and Industry 4.0 (Ind 4.0). Despite its advantages, SFC provisioning has significant challenges, including resource allocation, sequential VNF execution, extreme traffic management, and fulfilling end-to-end (E2E) latency limitations. These issues become more complex for dynamic applications with strict QoS requirements, such as CG, AR, and MIoT \cite{arda24}. 

Researchers utilize Deep Reinforcement Learning (DRL) algorithms for optimal VNFs' placements and SFC provisioning, as they perform well in decision-making and adapting to varying service demands \cite{liu24}\cite{onsu2024}. However, DRL models often rely on structured inputs and predefined policies, which can limit their adaptability in the presence of unexpected network conditions \cite{hoang23}. For instance, when a new SFC request arrives at a DC with insufficient computational resources, the DRL model applies its pre-trained policy based on historical data and structured inputs such as computational and storage capacity usage and SFC requests. A DRL model can struggle with this situation or any other unexpected situations such as sudden traffic surges or partial DC failures resulting in delays or suboptimal decisions as it requires multiple iterations to adjust its policy. Moreover, if a DRL agent takes the wrong action, it may require multiple training iterations to fix it, and if it keeps repeating the wrong action, it may reinforce the flawed policy rather than explore better alternatives, especially if the exploration-exploitation balance is not properly tuned. A weak reward function design may also fail to penalize incorrect decisions, allowing the model to continue misallocating resources without correction.

In contrast, Language Models (LMs) can complement DRL by analyzing both structured data (e.g., resource usage) and unstructured data (e.g., logs and  network configuration intent) in real-time \cite{boateng2024}. LMs can immediately detect bottlenecks, explain the issue (e.g., "The DC cannot handle the request due to insufficient computational power"), and suggest solutions, such as reallocating idle VNFs or rerouting the request to a less congested DC. Since LMs can interpret data without retraining for every new scenario, they provide rapid, informed recommendations and offer interactive Q/A capabilities for network state monitoring. This makes LMs particularly effective in handling dynamic and unpredictable network environments, enhancing DRL’s long-term optimization with real-time adaptability and responsiveness. This integration enhances E2E latency and service quality, allowing network operators to successfully meet the needs of dynamic applications, particularly those with variable workloads, strict latency requirements, and real-time adaptability needs, such as CG, AR, VS, and MIoT.

In our previous work \cite{arda24}, our focus was on VNF placement using DRL, where each VNF requires a certain amount of storage and computational power. Based on SFC requests and VNF resource requirements, the goal was to allocate resources optimally to maximize the number of SFC requests handled efficiently. In this paper, we extend that approach by integrating the network state conditions obtained after DRL actions into the LM for comprehensive network state monitoring. This integration enhances DRL-based VNF placement decisions by providing detailed insights into network dynamics. Given the importance of real-time adaptability and computational efficiency in network optimization, lightweight LMs offer more realistic alternative to large LMs \cite{lepagnol2024}. Large LMs demand huge computational resources and high-speed GPUs, but lightweight LMs can achieve comparable performance with substantially less hardware \cite{hassid2024}. Their higher inference speed and lower operational costs make them ideal for dynamic network environments requiring timely and efficient decision-making. 

In this paper, Bidirectional Encoder Representations from Transformers (BERT) model, a state-of-the-art lightweight LM known for its contextual understanding and ability to handle complex language tasks, is utilized and fine-tuned by Low-Rank Adaptation (LoRA). To evaluate efficiency and performance trade-offs, DistilBERT, a distilled version of BERT, is also used for comparison. The fine-tuned model is tested with different types of questions related to DC, VNF, and SFC information. This evaluation validates the LM's ability to assess resource usage and discover bottlenecks in order to make meaningful recommendations in the future. As a result, the system becomes more adaptable, allowing for proactive resource allocation, better scalability, and more efficient handling of SFC demands in dynamic network environments. 

The main contributions of this paper are as follows: 
\begin{enumerate}
\item A domain-specific dataset is created, derived from the DRL, incorporating current network state information as well as Q\&A related to SFCs, DCs, and VNFs. 
\item The dataset is fetched to BERT, which is fine-tuned using LoRA by modifying specific layers to achieve optimized performance tailored to the targeted application domain. DistilBERT is also applied to the dataset, and a comparative analysis is conducted between the models based on loss, confidence scores, and runtime efficiency.
\end{enumerate}

Based on our findings, BERT outperforms DistilBERT with lower test loss (0.28 vs. 0.36), higher confidence score (0.83 vs. 0.74), and a 4.87\% higher F1 score. However, this improvement comes at the cost of 46\% more processing time for BERT compared to DistilBERT. 

The literature review is provided in Section II, followed by the system model in Section III. Section IV discusses the language models under study, and performance evaluation is presented in Section V. We conclude in Section VI.

\section{Related Work}
Several strategies have been developed for SFC provisioning to optimize resource allocation while minimizing latency. Traditional optimization methods, such as Mixed Integer Linear Programming (MILP) \cite{pham020}, and heuristic approaches, such as Nearest Candidate Node Selection (NCNS)\cite{atinafu2024}, Fastest Candidate Node Selection (FCNS)\cite{atinafu2024}, and Priority-Based FCNS (PB-FCNS)\cite{atinafu2024}, utilize SDN and queueing models to efficiently assign computing nodes \cite{atinafu2024}. While these solutions reduce latency and increase service success rates, they rely on predefined decision criteria and are not adaptable to dynamic network conditions. 

RL can outperfrom heuristic approaches in SFC provisioning by continuously learning and optimizing from interactions with the environment rather than relying on predefined rules.
Thinh et al. \cite{tran2024} has proposed a DRL-based framework for SFC provisioning that optimizes VNF embedding and routing while meeting time and resource constraints. They have compared Deep Q-learning (DQL) and Advantage Actor-Critic (A2C), demonstrating that both achieve over 95\% service request acceptance along with high network throughput, performing similarly to optimization-based methods but with considerably shorter execution times. 
In another study  by Onsu et al." \cite{arda24}, a DRL-based SFC provisioning approach has been proposed that offers reconfigurability of networks without the need to retrain the model in different network setups. Moreover, the advanced DNN architecture for their DRL model includes an attention layer to optimize resource utilization \cite{arda24}. 

Although DRL is effective for SFC provisioning, it has drawbacks such as  slow adaptation to unexpected changes, and high training costs while correcting an agent wrong decisions. Recently, a few studies have explored the use of LMs for SFC provisioning.
Van et al.\cite{VAN24} have utilized LMs for intent-based NFV configuration, translating natural language intents into JSON templates that include attributes such as VNF name, type (e.g., firewall, IDS), computational requirements (CPUs, memory, storage), and actions (create, update, destroy). The proposed system has been integrated with NI-testbed, an AI-driven NFV lifecycle management framework, to automate VNF and SFC deployment \cite{VAN24}. However, the proposed system lacks adaptive learning capabilities, meaning that decision-making does not improve over time. If an SFC request is suboptimal, the system does not learn from previous mistakes to make better decisions in the future. Instead, it automatically converts users requests to JSON without considering whether they are ideal for the present network conditions. While the intent is not deployable, it only warns the administrator of errors rather than optimizing or modifying configurations on its own.
Additionally, the research does not address whether a VNF is idle or  whether the system has enough resources to handle all requests. This information is crucial for ensuring optimal resource usage, preventing over- or under-provisioning, and preserving Quality of Service (QoS) while reducing operational costs.

Li et al.\cite{Li2025} have proposed an LLM-assisted  SFC optimization that uses LLMs to dynamically create heuristic functions for solving optimization problems and a Non-dominated Sorting Genetic Algorithm (NSGA) to refine the heuristic functions generated by the LLMs. However, the LLM-generated heuristics are fixed until manually changed, which prevents automatic adjustments to unexpected network changes. While the proposed model optimizes SFC deployment over multiple iterations, it does not  store previous network states or adapt in real time. If conditions change after optimization, the system must be restarted from scratch, which limits continuous learning and real-time adaptability.
 \begin{figure*}[!t]
        \centering
        \includegraphics[width = 0.8\textwidth, trim=0cm 0cm 0cm 0cm,clip]{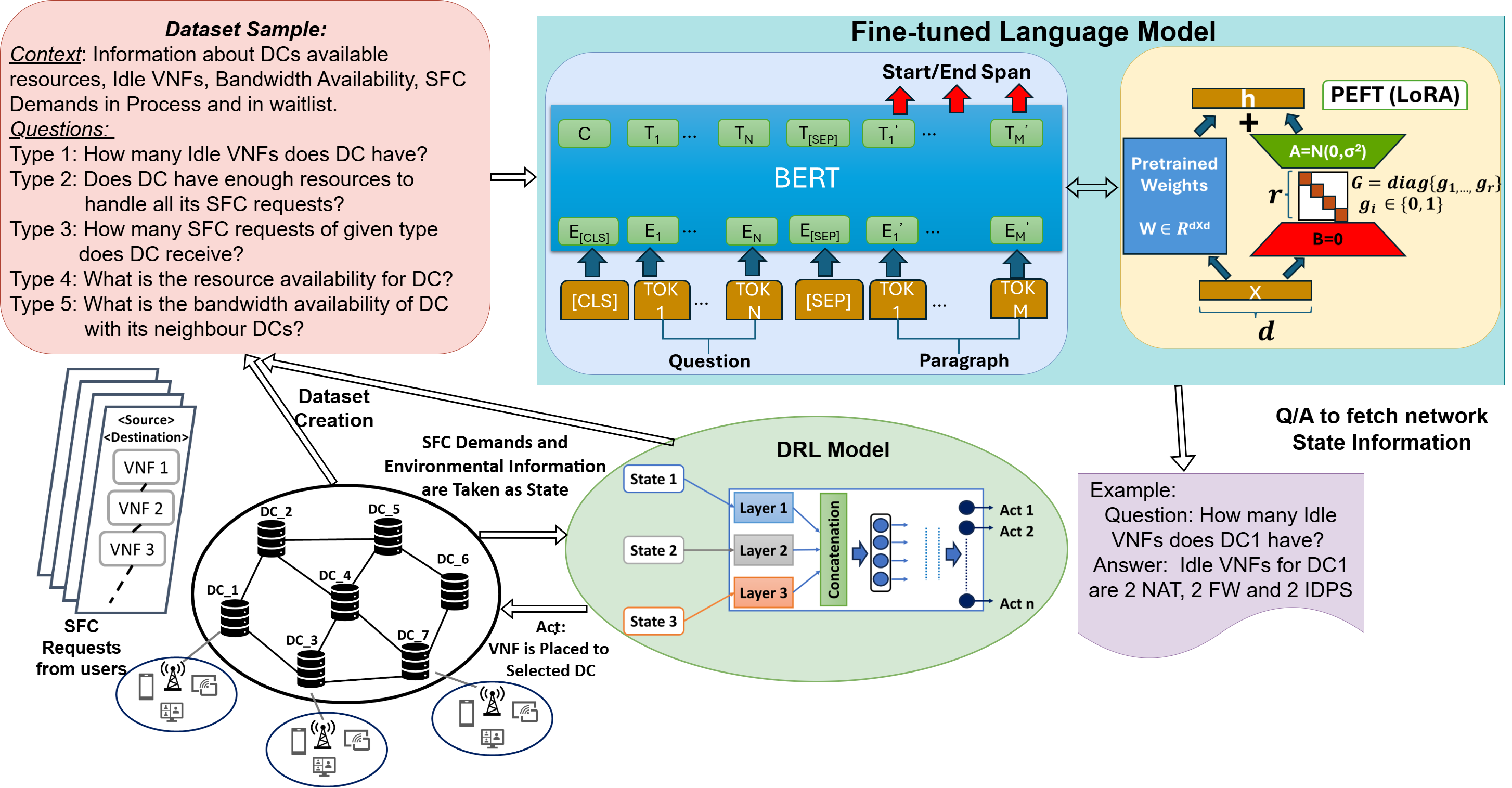}
        \caption{System Architecture for LM-Assisted Network State Monitoring with DRL-Driven SFC Provisioning.}
        \label{fig:system} 
\end{figure*}
Our proposed methodology, based on our previous work \cite{arda24}, integrates  LMs to monitor the network's current state that can direct DRL to improve SFC provisioning and VNF placement. By leveraging DRL-learned patterns and following LM's instructions the system can dynamically optimize SFC deployment, ensuring efficient resource utilization and improved QoS. Additionally, this approach adjusts to changing network conditions without requiring manual heuristic updates or re-optimization from scratch. Furthermore, by using DRL-generated outputs, SFC's and DCs'  information as input for LM, our model improves system interpretability and enables a more detailed analysis of SFC configurations, resource allocation, and VNF state for future requests.

\section{System Model}
In our previous study \cite{arda24}, the objective was to maximize accommodated SFC requests while adhering to infrastructure constraints by using DRL. Different SFC requests have varying resource demands, latency requirements, and VNF sequences, making placement decisions highly dependent on the specific characteristics of each service. The types of SFCs used in this work are provided in detail in \tablename\hspace{0.1pt}~\ref{tab:sfc_characteristic}. DRL demonstrated effective resource utilization and network performance optimization utilizing learned policies. However, it responds slowly to unexpected changes, and if it encounters incorrect or suboptimal decisions, it may require extensive retraining or fine-tuning to correct its policy, leading to delays in adaptation. In contrast, LMs can generate fast real-time context-aware VNF placement instructions, making DRL actions more responsive and versatile in dynamic network environments.

In our current research, we utilize LM for network state monitoring that would instruct the DRL model to improve VNF placement decisions.  LMs can assist with learning and interpreting existing network conditions, generating useful insights that can be applied to future requests. By processing current information on bandwidth modifications, latency variations, and resource availability, LMs can generate comprehensive understanding of the network’s state. This data can be utilized not only to make fast decisions, but also to recommend the best placement of VNFs for new incoming requests depending on resource requirements and the current network state. As a result, the system becomes more adaptive, allowing for proactive resource allocation, increased scalability, and more efficient processing of SFC requests. As shown in \figurename\hspace{0.1pt}\ref{fig:system} we generate a domain-specific dataset based on the network state information and action updates obtained from the DRL, and feed these context and 5 types of Q/A to a pre-trained BERT that has been fine-tuned for our specific dataset using LoRA. These questions extract key information such as the number of idle VNFs available in a DC, whether a DC has enough resources to process all its SFC requests, and the number of SFC requests of a specific type received by a DC. Furthermore,  a DC's total resource availability and its bandwidth availability in comparison to neighboring DCs are analyzed. The dataset also addresses critical components of DC performance, including computational power, storage capacity, and network connectivity. Additionally, it provides insights into DC connections and operational status, such as bandwidth availability between neighboring DCs and the number of idle VNFs. These structured queries play a crucial role in efficient monitoring and managing  resources for optimal SFC provisioning.

\begin{table} 
\centering
\caption{SFC characteristics \cite{table}}\fontsize{6.5}{7.7}\selectfont
\begin{tabular}{|p{1.4cm}|p{1.6cm}|p{1cm}|p{1cm}| p{1 cm}|} 
 \hline
 \textbf{SFC Request}&\textbf{VNF Sequence} &\textbf{Bandwith (Mbps)} &\textbf{E2E delay (msec)} &\textbf{Request Bundle} \\  
 \hline
  CG & NAT-FW-VOC\break-WO-IDPS  & 4 & 80  & [40-55] \\
  \hline
  AR & NAT-FW-TM\break-VOC-IDPS & 100 & 10  & [1-4] \\
  \hline
  VoIP & NAT-FW-TM\break-FW-NAT & 0.064 & 100 & [100-200] \\
  \hline
  VS & NAT-FW-TM\break-VOC-IDPS & 4  & 100  & [50-100] \\
  \hline
  MIoT & NAT-FW-IDPS  & [1-50] & 5 & [10-15] \\
  \hline
  Ind 4.0 & NAT-FW & 70 & 8 & [1-4] \\
 \hline
\end{tabular}
\label{tab:sfc_characteristic}
\end{table}

\section{Language models under study}
In this study, BERT and DistilBERT are utilized and fine-tuned with LoRA to improve efficiency for our domain-specific dataset. BERT is chosen for its outstanding contextual understanding, and compared with DistilBERT, a lighter and faster counterpart. Compared to larger LMs, lightweight LMs provide faster inference, lesser resource requirements, and easier fine-tuning, making them ideal for dynamic network conditions \cite{lepagnol2024}.

BERT is a transformer-based model that has been pre-trained to perform natural language understanding tasks such as question answering, text classification, and language inference. Unlike traditional models, BERT employs a bidirectional attention mechanism that allows it to capture contextual information from both prior and following tokens in a phrase. This makes BERT particularly useful for tasks that require an extensive understanding of the context. However, fine-tuning BERT for domain-specific tasks frequently requires updating millions of parameters, which can be computationally intensive. To solve this, LoRA offers a more efficient solution by freezing the majority of the model's parameters and inserting trainable low-rank matrices into certain layers. Low-rank matrices adjust self-attention outputs with fewer parameters, preserving BERT's expressiveness while improving fine-tuning efficiency for domain-specific tasks.

In this study, the input data for BERT contains contexts, questions, and answers about SFCs, VNFs, and DCs. BERT uses WordPiece tokenization to handle out-of-vocabulary terms by splitting them into subword units. To improve input representation and reduce token usage, we modify the Hugging Face AutoTokenizer by explicitly including domain-specific words to the tokenier. These domain-specific words, including "CG", "DC", "NAT" (Network Address Translation), "FW" (Firewall), "VOC" (Video Optimization Controller), "WO" (WAN Optimizer), "IDPS" (Intrusion Detection and Prevention System), "VNF", "E2E", "MBPS", "BW" (bandwidth) and "Ind40", allow the tokenizer to represent specialized words as single tokens rather than multiple sub-tokens, ensuring efficient utilization of the available max length. This strategy not only increases encoding efficiency by preserving critical context space, but also improves BERT's knowledge of domain-specific concepts. 

LoRA is applied on specific layers of the BERT model, specifically the query and value projection matrices in the self-attention mechanism of each transformer layer. This enables the model to be fine-tuned with fewer trainable parameters by using low-rank decomposition matrices to modify the attention projections without updating the whole weight matrices. Once the training is completed, the model is evaluated by giving unseen descriptions and questions and comparing the predicted answers to the dataset's ground truth answers. This comparison enables us to evaluate the model's ability to reliably extract relevant information and confirm that the Q/A framework works well in real-world scenarios. 

Additionally, DistilBERT is employed on the dataset and fine-tuned with LoRA.  DistilBERT is a lighter and faster version of BERT, designed to retain performance close to that of BERT while being more efficient.  It achieves this using knowledge distillation, a method in which a smaller model (DistilBERT) learns from a bigger, pretrained model (BERT) while retaining crucial language understanding capabilities \cite{sanh2019}. DistilBERT decreases computational cost and inference delay, making it ideal for real-time applications, low-resource situations, and tasks that require quick processing without compromising too much accuracy. The balance between efficiency and performance makes DistilBERT an excellent choice for network optimization and VNF placement, where fast and adaptive decision-making is essential.

\section{Performance Analysis}
The experiments were carried out on a system equipped with NVIDIA A100-PCIE-40GB GPUs, each of which features 40GB of memory. The JSON dataset includes 1920 sets of context, questions, and answers. It is divided into three parts: 75\% for training, 12.5\% for validation, and 12.5\% for testing with a learning rate of 2e-5 and a batch size of 4.
We use BERT-base-uncased model, with a maximum sequence length of 512 tokens. This model consists of 12 levels (transformer blocks), each with 768 hidden dimensions, for a total of 110 million parameters.  Additionally, we train DistilBERT-base-uncased model with a maximum sequence length of 512 tokens, 6 levels (transformer blocks each with 768 hidden dimensions), and for a total of 66 million parameters. Using knowledge distillation during training, DistilBERT effectively compresses the knowledge of the full BERT model into a smaller architecture \cite{sanh2019}. 
For LoRA, we configured r=8, representing the rank of the low-rank update matrices, and \(\alpha\)=16, as the scaling factor for these updates. 
A dropout rate of 0.08 was applied to regularize the LoRA layers, targeting specific modules like "query" and "value" in the attention mechanism for BERT and q\_lin (Query Projection Layer) and k\_lin (Key Projection Layer) for DistilBERT. All of the configuration parameters are summarized in \tablename\hspace{0.1pt}~\ref{tab:training_params}, which have been chosen empirically. 

\begin{table}[!t]
    \centering
    \begin{tabular}{|l|c|}
        \hline
        \textbf{Parameter} & \textbf{Value} \\
        \hline
        Learning Rate & 2e-5 \\
        Batch Size & 4 \\
        Max Length of Tokens & 512 \\
        LoRA Rank (r) & 8 \\
        LoRA Scaling Factor ($\alpha$) & 16 \\
        \hline
    \end{tabular}
    \caption{ LoRA Configuration Parameters}
    \label{tab:training_params}
\end{table}

\begin{figure}[!hbt]
    \centering
    \begin{subfigure}[b]{0.38\textwidth}
        \centering
        \includegraphics[width=\textwidth, trim=0cm 0cm 0cm 0cm, clip]{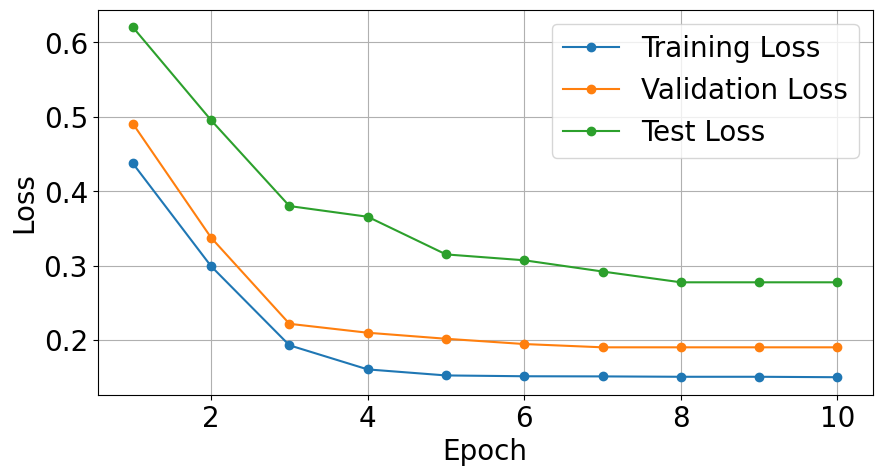}
        \caption{Loss for BERT}
        \label{fig:loss}
    \end{subfigure}
    \begin{subfigure}[b]{0.38\textwidth}
        \centering
        \includegraphics[width=\textwidth, trim=0cm 0cm 0cm 0cm, clip]{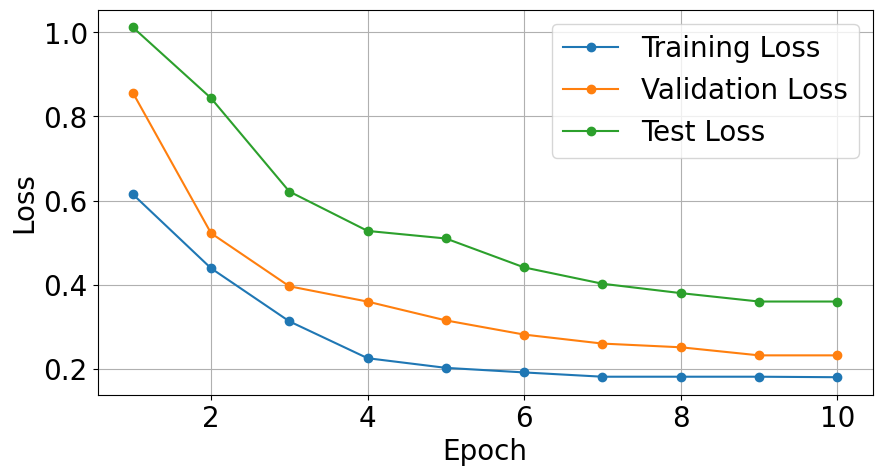}
        \caption{Loss for DistilBERT}
        \label{fig:loss2}
    \end{subfigure}
    \caption{Training, testing, and validation loss for BERT and DistilBERT}
    \label{fig:metrics}
\end{figure}
\figurename\hspace{0.1pt}\ref{fig:loss} illustrates  the BERT model's training, validation, and test loss over ten epochs. Initially, all three losses are high, with training and validation losses decrease steadily throughout the first few epochs, demonstrating effective learning during the early phases of training. Test loss begins higher but rapidly decreases as compared to training and validation losses, indicating that the model generalizes moderately well to unseen data. Compared to the BERT model, the DistilBERT model, depicted in \figurename\hspace{0.1pt}\ref{fig:loss2},  begins with higher initial losses across all three metrics but then rapidly decreases during the early epochs. By the eighth epoch, the losses have stabilized, with the training loss being the lowest, followed by the validation and test losses. When comparing BERT to DistilBERT, the trade-off between performance and efficiency is evident.

We also evaluate the models based on confidence score which measures the model's certainty about its predicted answer. It is calculated by converting logits for the start and end positions into probabilities using the softmax function. The confidence score for a span is the product of the start and end probabilities, and the span with the highest score is selected as the prediction. The average confidence is obtained by calculating the confidence score for each prediction and taking the mean across all predictions.

\begin{figure}[!t]
        \centering
        \includegraphics[width = 0.4\textwidth, trim=0cm 0cm 0cm 0cm,clip]{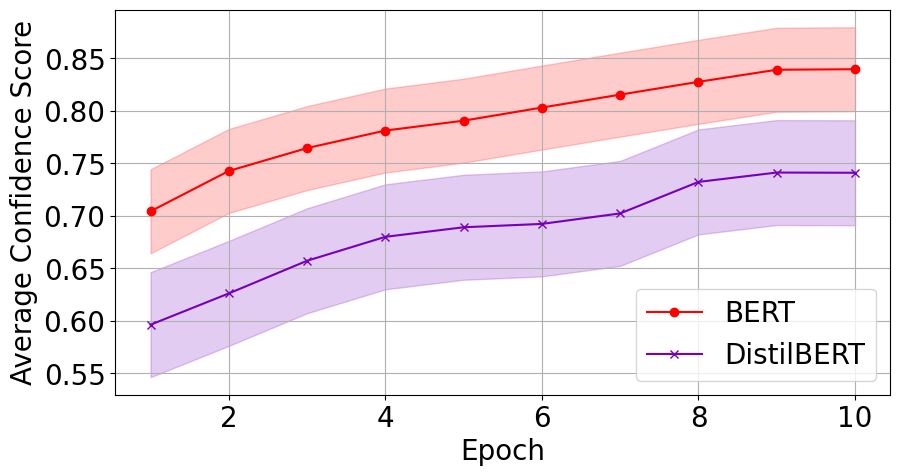}
        \caption{ Confidence score for BERT and DistilBERT }
        \label{fig:confidence} 
\end{figure}

 \figurename\hspace{0.1pt}\ref{fig:confidence} illustrates the average confidence scores of predictions made by BERT and DistilBERT models over 10 epochs, along with their respective confidence intervals. Based on \figurename\hspace{0.1pt}\ref{fig:confidence}, BERT constantly has higher confidence scores, beginning at around 0.7 in the first epoch and progressively growing to around 0.83 by the 10th epoch, showing constant improvement and strong learning. DistilBERT, on the other hand, starts with lower confidence score of about 0.59 and rises to around 0.74 by the last epoch, indicating a tradeoff between performance and efficiency due to its lighter architecture. The confidence intervals clearly demonstrate the differences in prediction consistency between the two models. BERT has narrower confidence intervals, which means that its predictions are more consistent and less varied among samples. In comparison, DistilBERT has wider confidence intervals, indicating that its predictions are less consistent and more variable.

\begin{figure}[!t]
        \centering
        \includegraphics[width = 0.38\textwidth, trim=0cm 0cm 0cm 0cm,clip]{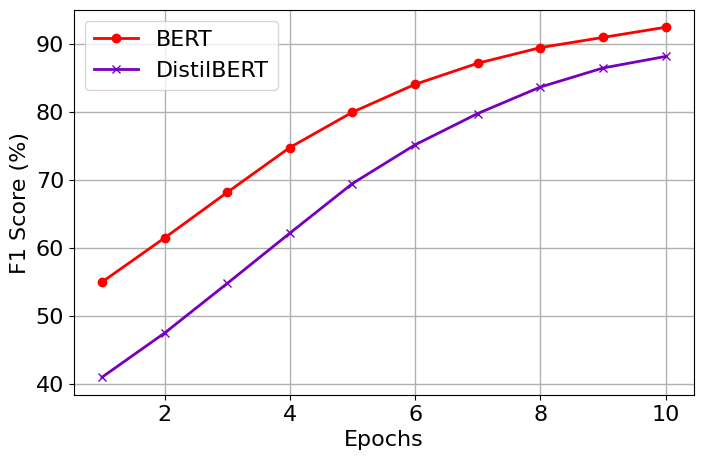}
        \caption{ F1 score (\%) for BERT and distilBERT}
        \label{fig:f1score} 
\end{figure}

 F1 score is a crucial metric for evaluating the model's ability to correctly extract answer spans from a given context. It is calculated as the harmonic mean of precision and recall, where precision is the proportion of correctly predicted words out of the total predicted words and recall is the proportion of correctly predicted words out of the total words in the ground truth answer. 
F1 score for BERT and distilBERT are provided in \figurename\hspace{0.1pt}~\ref{fig:f1score}, where BERT achieves an F1 score of 92.5\%, while DistilBERT reaches 88.2\% in the last epoch. In other words, BERT performs 4.87\% better than DistilBERT in terms of F1 score.

\begin{figure}[!t]
        \centering
        \includegraphics[width = 0.38\textwidth, trim=0cm 0cm 0cm 0cm,clip]{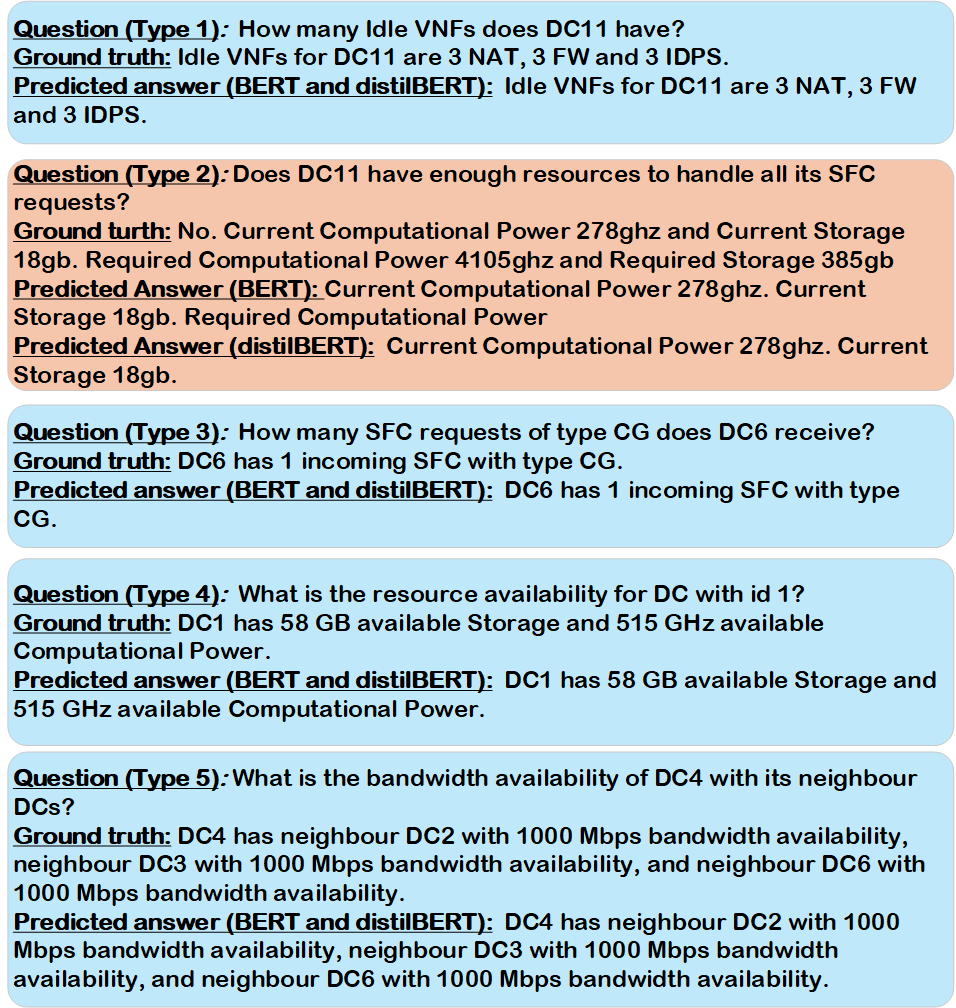}
        \caption{ Samples of five types of questions, their ground truth answers, and predicted answers}
        \label{fig:q/a} 
\end{figure}

Regarding the ability to predict answers, as depicted in \figurename\hspace{0.1pt}\ref{fig:q/a} both BERT and DistilBERT successfully can predict answers for question types 1, 3, 4, and 5, indicating their ability to extract relevant information from the dataset. However, both models struggle with question type 2, that evaluates whether an SFC can manage all of its requests. The challenge arises because answering the question type 2 requires information about the current computational capability (which is given in the context), calculating required resource demand for all VNFs by summing their computational demands, and comparing current and required computational resources, which is a process that goes beyond typical text-based reasoning. BERT and DistilBERT are pre-trained language models that are primarily intended for natural language interpretation rather than mathematical computations or real-time state-based decision-making, hence they cannot do numerical calculations \cite{rogers2020}. As a result, their answers to such questions are often incorrect or incomplete.

In terms of total processing time for our dataset, BERT took 4 hours, 8 minutes, and 38.02 seconds, whereas DistilBERT completed the process in 2 hours, 14 minutes, and 46.55 seconds, making it approximately 46\% faster.
While BERT delivers higher and more consistent confidence score with less loss, it needs more computational resources. On the other hand, DistilBERT, while slightly less confident and consistent with slightly bigger loss, can be an efficient alternative that balances acceptable performance with lower computational demands.
This makes it ideal for scenarios where computational efficiency has priority over minor improvements in performance.

\section{Conclusion}
In this work, the network state information, specifically the final VNF allocations determined by the DRL, SFCs, and DCs' information, are fetched to the LM, enabling it to interpret and respond to queries regarding the state of SFCs, DCs, and VNFs. Inquiring about the state of SFCs, DCs, and VNFs gives insight into current resource utilization and bottlenecks, allowing better planning and response to future requests. After fine-tuning LMs using LoRA to improve adaptability to our domain-specific dataset, BERT outperformed DistilBERT in terms of lower losses and higher confidence scores, with the confidence score for BERT converging to 0.83 compared to 0.74 for DistilBERT. However, a better performance of BERT comes at the cost of a 46\% longer processing time compared to DistilBERT. In our ongoing work, large language models such as GPT-3.5 are applied to the dataset to evaluate whether using a larger model leads to better answers or if BERT and DistilBERT are already sufficient for this dataset. We are also planning to utilize the LMs' analysis to instruct the DRL model to improve the network state and optimize SFC request handling dynamically. Moreover, this technique will be integrated into an intent-driven management framework, allowing more autonomous and context-aware decision-making in network orchestration.
 

\section*{Acknowledgment}
This work is supported by the Natural Sciences and Engineering
Research Council of Canada (NSERC) Alliance Program, MITACS
Accelerate Program, and NSERC CREATE TRAVERSAL program.
\bibliographystyle{IEEEtran}

\end{document}